# Local fluid pressure perturbations inside faults matter in unexpectedly large injection-triggered earthquakes

Yinlin Ji[1], Hannes Hofmann[1], Kang Duan[2], Arno Zang[1]

[1] *GFZ German Research Centre for Geosciences, 14473 Potsdam, Germany*

[2] *School of Civil Engineering, Shandong University, 250061 Jinan, China*

*Corresponding author: Yinlin Ji (yinlinji@gfz-potsdam.de)*

*Co-authors:* Hannes Hofmann (hannes.hofmann@gfz-potsdam.de); Kang Duan (kang.duan@sdu.edu.cn); Arno Zang (zang@gfz-potsdam.de)

Anticipating the maximum magnitude of injection-triggered earthquakes is highly valuable for the safe and efficient exploitation of geoenergies. The recent work by Li et al.[1] reached the conclusion that unexpectedly large injection-triggered earthquakes are primarily caused by large pre-existing critical shear stresses on seismogenic faults. Also of great interest is the proposal of the ratio of fault slip to dilation as an index to anticipate the fault rupture energetics. However, their fluid injection experiments were conducted under fully drained conditions, where the fluid pressure distribution on the fault plane is always uniform. As already pointed out by the authors, local fluid pressure perturbations inside faults under locally undrained conditions also have the potential to cause unexpectedly large seismic events. Here we add more datasets and possible explanations for this viewpoint, which has not been explored in such detail by Li et al.[1].

Since the earthquake magnitude is proportional to the fault size[2], unexpectedly large injection-triggered earthquakes are produced on extraordinarily large faults. We argue that a local fluid pressure perturbation inside a fault is important in the case of large magnitude earthquakes, because the injected fluid tends to localize in a fault segment adjacent to the injection well, forming an overpressure pocket[3] (Fig. 1a). Local fluid pressurization on faults can be reproduced in the laboratory by regulating the injection rate and the pre-existing stresses on a fault with various roughnesses[4-7] (Fig. 1b). To quantify the fluid pressure heterogeneity on the fault plane, we define the fluid overpressure ratio ($R=P_{inj}/P_{MC}$) as the ratio between the fluid pressure at the onset of fault failure measured at the injection point ($P_{inj}$) and that predicted by the Mohr-Coulomb failure criterion ($P_{MC}$) combined with Terzaghi's effective stress law[5] (Fig. 1b). When the fluid pressure distribution is uniform on the fault plane, the failure criterion is expected to hold and the fluid overpressure ratio is unity as in Li et al.[1], and this ratio increases with more heterogeneous fluid pressure distribution. The fluid overpressure ratio increases with faster injection rate, higher initial normal stress, and higher shear stress as exhibited in Fig. 2a, which is plotted based on the data extracted from Passelègue et al.[7]. These trends are understandable, because the fluid pressure distribution on the fault plane depends on the balance between fluid injection and diffusion rates: a faster rate of fluid injection promotes the accumulation of fluid around the injection point; a higher initial normal stress squeezes the fault and reduces the fault permeability, making it difficult for the fluid to diffuse along the fault; a higher shear stress means a more critically stressed fault and allows shorter time for the fluid to diffuse before the onset of fault reactivation.

As shown in Fig. 2b and 2c, the fluid overpressure ratio is a key parameter influencing the injection-induced seismic hazard, measured by the slip rate at fault reactivation[7] (Passelègue et al. 2018) and the maximum seismic moment[4,5]. The maximum seismic moment here is also a proxy for deformation moment magnitude estimated from the laboratory-derived shear modulus, fault area and shear displacement jump. This could be explained by the fact that the local fluid perturbation acts as a point load to perturb the stability of a prestressed fault by transferring shear stress beyond the overpressure pocket[8,9] (Fig. 1, failure mechanism 2). The intensity of the perturbation increases with larger fluid overpressure ratio characterized by a higher injection pressure at fault failure[6]. As demonstrated by Galis et al.[10], the capability of rupture propagation along a fault scales with the intensity of perturbation. Thus, a larger fluid overpressure ratio at fault failure results in a faster slip rate and a larger magnitude of maximum seismic moment. We also evaluate the relationship between the maximum seismic moment and total injected volume in Fig. 2c. The three test series refer to the fluid injection experiments on prestressed faults each under the same stress state but various injection scenarios. Despite the scatter of data points, the maximum seismic moment generally increases with larger total injected volume roughly consistent with the trend predicted by McGarr[11]. In each test series, both the maximum seismic moment and total injected volume increases with higher fluid overpressure ratio, which also tends to elevate the data points above the McGarr boundary, similar to the effect of elevating shear stress as reported in Li et al.[1]. In addition, the ratio of fault slip to dilation is complicated by the heterogeneous fluid pressure distribution and the associated local fault slip and dilation[3,8,12], making it less effective as a proxy for the impending fault rupture initiation and propagation.

We emphasize that Li et al.[1] presented an excellent piece of work to further constrain the maximum seismic moment induced by fluid injection based on the pre-existing shear stress and to anticipate the seismic potential from the ratio of fracture slip to dilation. Considering the large fault size necessary for accommodating large seismic events, local fluid perturbations inside faults discussed in this manuscript serve as a complementary and coupled process with high pre-existing shear stresses for triggering unexpectedly large earthquakes (Fig. 1). To further constrain the maximum seismic moment, one should incorporate more details of the architecture of natural fault zones, like segmented faults with fracture walls of different roughnesses, both in laboratory- and mine-scale in-situ experiments to assess the mutual impact of fluid pressure heterogeneity and critical shear stress perturbation on the injection-induced earthquake rupture processes.

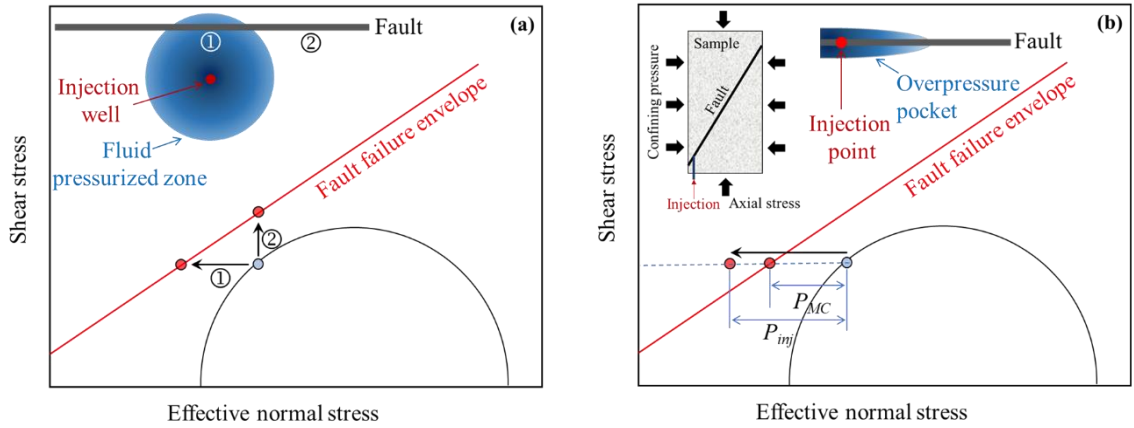

**Fig. 1** Local fluid pressure perturbations inside a **(a)** field and **(b)** lab fault prone to failure. The circles filled with blue and red colours represent the stress states on fault segments before and upon reactivation. **(a)** There are two coupled mechanisms (numbered as 1 and 2) of fault reactivation subject to local pressure perturbation. The first failure mechanism is to reduce the shear strength of the fault segment by elevating fluid pressure within the fluid pressurized zone, as explored in Li et al.[1]. The second failure mechanism is to elevate the shear stress on the fault segment beyond the fluid pressurized zone. Fault reactivation can occur at several locations by different mechanisms, complicating the predictions of rupture nucleus and event magnitude. **(b)** The lab fluid injection experiments on prestressed faults were conducted using the triaxial shear-flow setup and the fluid is introduced locally to the fault through the injection point. The formation of a pressure pocket on a fault is favoured by a higher normal stress, shear stress and injection rate. $P_{MC}$ and $P_{inj}$ are the fluid pressures at the onset of fault reactivation predicted by the failure criterion and measured at the injection point, respectively. They are equal only when the fluid pressure distributes uniformly on the fault, otherwise $P_{MC}$ is large than $P_{inj}$.

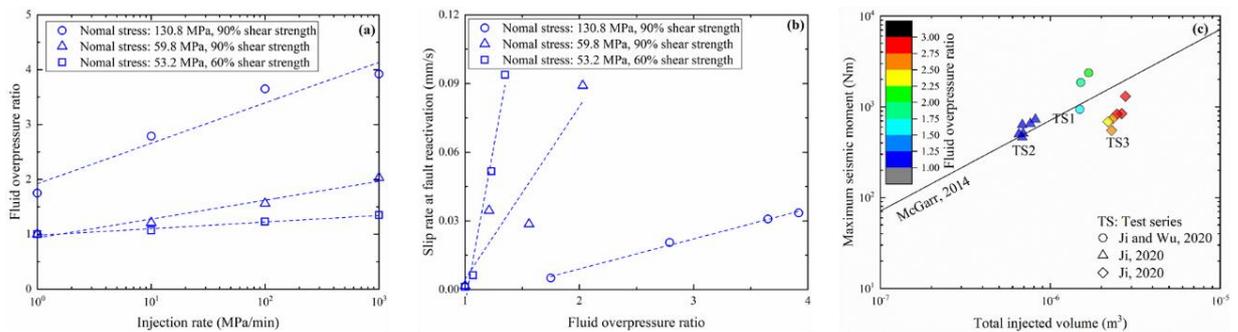

**Fig. 2 (a)** Fluid overpressure ratio as a function of injection rate[7]. **(b)** Slip rate at fault reactivation as a function of fluid overpressure ratio[7]. **(c)** Maximum seismic moment as a function of total injected volume with McGarr boundary[11] and experimental datasets[4, 5]. Each test series was conducted under the same stress state with various injection scenarios. Symbols filled with colours from cold to hot represent the increasing fluid overpressure ratio.

### Acknowledgements

This work has been supported by the Helmholtz Association's Initiative and Networking Fund for the Helmholtz Young Investigator Group ARES (contract number VH-NG-1516).

## Author contributions

Y．J．conceived the study, performed data collection and analysis, and prepared the manuscript and figures. H. H., K. D., and A. Z. reviewed and edited the writing. All authors discussed, reviewed, and edited the manuscript and ultimately agreed to the submitted version of the manuscript.

## Competing interests

The authors declare no competing interests.